\documentstyle[12pt]{article}
\begin{document}
\begin{flushright}
hep-th/0412333\\
SNB/December/2004
\end{flushright}
\vskip 2.5cm
\begin{center}
{\bf \Large { Time-Space Noncommutativity And Symmetries\\
For A Massive Relativistic Particle}}

\vskip 2.5cm

{\bf R.P.Malik}
\footnote{ E-mail address: malik@boson.bose.res.in  }\\
{\it S. N. Bose National Centre for Basic Sciences,} \\
{\it Block-JD, Sector-III, Salt Lake, Calcutta-700 098, India} \\

\vskip 2cm

\end{center}

\noindent
{\bf Abstract}: 
We show the existence of a time-space noncommutativity
(NC) for  the physical system of a massive relativistic particle
by exploiting the underlying symmetry properties
of this system. The space-space NC is eliminated by the
consideration of the exact symmetry properties and their consistency with the
equations of motion for the
above system. The symmetry corresponding to the noncommutative
geometry turns out to be the special case of the gauge symmetry
such that the mass parameter of the above system becomes
noncommutative with the space and time variables. 
The possible deformations of the gauge
algebra between the spacetime variables and the angular momenta
are discussed in detail. These modifications owe their origin to the
NC of the mass parameter with the space and time variables. 
The cohomological origin for the above NC is addressed in the language of 
the off-shell nilpotent Becchi-Rouet-Stora-Tyutin
(BRST) symmetry transformations. \\

\baselineskip=16pt

\vskip .7cm



\newpage

\noindent
{\bf 1 Introduction}\\

\noindent
The subject of noncommutative field theories has been a topic of
intensive research activities during the past few years. Despite the fact 
that the idea of noncommutativity (NC) in the spacetime structure
has a long history [1,2], the recent
upsurge of interest in the NC (and the corresponding field theories 
constructed on the noncommutative spacetime) has
been spurred due to its very clear and cogent appearance in the context
of string theories, $D$-branes and $M$-theories. To be more precise , the end 
points of the strings, trapped on the $D$-branes, turn out to be 
noncommutative in the presence of a  background two-form gauge field [3,4].
It has been argued, furthermore, that the string dynamics
could be shown to be equivalent to the minimally coupled gauge 
field theory defined on a noncommutative space [5]. From a distinctly
different point of view, a careful study of the
quantum gravity and black hole physics entails
upon the spacetime structure to become noncommutative in nature [6,7]. In
other words, an attempt to unify the key concepts of
quantum mechanics and  gravity
(which is a benchmark of theoretical physics at an energy scale comparable to
the Planck energy)
requires, by various considerations, a spacetime structure where
the $D$-dimensional spacetime variables $x_\mu$ $(\mu = 0, 1, 2...D-1)$
become noncommutative (i.e. $[x_\mu, x_\nu ] \neq 0$) leading to 
an uncertainty relation between the two of them.

It is evident from the above discussions
that the impact of the NC in spacetime structure
would manifest itself only
at a very high energy scale. However, it might also be possible to
physically test its very existence through the low-energy effective
actions, obtained in some specific limits, from the actual 
high energy theory. This is why
a few experiments have been suggested in the literature
to observe the impact of these NCs on some physically interesting
quantities. It has also been argued that only the lowest order
quantum mechanical effects are good enough to shed some light on the
spacetime NC in the context of celebrated Aharonov-Bohm type
of experiments and/or synchrotron radiation studies [8-10]. However, so far,
there have not been any successful observations of these effects. There
is an alternative way to theoretically predict some consequences
of the spacetime NC on some physically interesting quantities. In this
approach, one can exploit the basic 
concepts of the noncommutative geometry (as well as
the  corresponding noncommutative field theories)
and construct the low energy effective actions. In this category, one
can mention a couple of attempts that are basically connected
with the noncommutative Chern-Simons theory [11] and the
noncommutative standard model of unification [12]. The physical consequences
of these studies are, however, yet to be tested experimentally.

The reparametrization invariant models of the free as well as
interacting relativistic (super)particles are at the
heart of the modern developments in the (super)string theories and 
(super)gravity theories. These models are also found to be endowed
with the first-class constraints in the language of the Dirac's
prescription for the classification of constraints [13,14]. As a
consequence, the above models respect a 
set of local gauge symmetry transformations that is generated by the 
first-class constraints themselves. Thus, the above models clearly
possess a whole host of interesting mathematical 
as well as physical structures. The purpose of our present paper
is to study the model of the free massive relativistic particle and
to demonstrate the existence of a specific variety of
time-space NC in the spacetime structure (see, e.g., [15-17] for details)
by exploiting the continuous symmetry properties of the Lagrangians for the
above system. In particular, we lay stress on the
standard- as well as non-standard gauge type of symmetry transformations
(and their consistency with the equations of motion)
which imply  the presence of commutativity and NC in the theory. It is 
worthwhile to mention that a whole range of studies, connected with
the unitary quantum mechanics and their physical
consequences [15-17], have been performed based on the above
type of NC. Thus, the above time-space NC is physically very interesting.

It is pertinent to point out that, in a couple of papers [18,19],
the reparametrization invariant models and their mechanics have been studied
where the equivalence of the NC and commutativity in the spacetime structure
has been demonstrated in the language of the Dirac bracket formalism
for two gauge choices. The deformation of the Poincar{\' e} algebra
and the algebra between the spacetime variables and the angular momenta
has been obtained due to the presence of the NC in spacetime. For the massless
relativistic particle (that is endowed with more symmetry properties
than its massive counterpart), an extension of the total
conformal algebra has been shown to exist due to the NC in spacetime [20].
The latter emerges due to the presence of
a {\it new} scale type of symmetry (that is different from
the global scale symmetry of the usual conformal group of spacetime 
transformations). The dynamical implications of the above NC have been
studied thoroughly in the language of the symplectic structure and
Poisson bracket formalism [21]. In a very recent paper [22], a toy model of 
the reparametrization invariant free non-relativistic particle 
(see, e.g., [19])
has been studied in detail where the equivalence between the NC and
commutativity has been demonstrated in the language of
the standard- and non-standard gauge type of symmetry transformations
for the Lagrangian of the system. As it turns out, the symmetry corresponding
to the noncommutative geometry entails upon the mass parameter of this
model to become noncommutative in nature with the space variable {\it alone}.

In our present paper, we generalize our earlier idea [22], to the
reparametrization invariant model of a free massive relativistic particle
and show the existence of the NC in spacetime structure by exploiting
the continuous symmetry properties of the Lagrangian for the system.
It should be emphasized that we obtain a unique symmetry transformation
(see, e.g., (3.10) below) from the standard gauge symmetry transformations
(corresponding to a commutative geometry) as well as from the
non-standard gauge-type of symmetry transformations (corresponding to
a noncommutative geometry). The important point, to be noted, is the fact
that one can {\it not} guess the unique symmetry transformations
(see, e,g, (3.10) below) from the standard gauge symmetry transformations.
Rather, it is logically derived from the non-standard gauge type of 
transformations (corresponding to a noncommutative geometry) by requiring
the consistency between the non-standard symmetry transformations and the 
equations of motion derived from the 
equivalent Lagrangians for the system. In
the above consistency requirement, the expressions for the 
canonical momenta, obtained from the different (but equivalent) Lagrangians,
play very important role. It is clear that the unique symmetry
transformation (see, e.g., (3.10) below) entails upon (i) the spacetime
to become commutative from one point of view (i.e. standard
gauge transformations), and (ii) the spacetime to become noncommutative
from another point of view (i.e. the non-standard gauge type transformations).
This establishes a couple of important
results in one stroke. First,
it demonstrates the equivalence of the commutativity and NC in the theory
which agrees with such an observation made in the language of the Dirac
bracket formalism (see, e.g., [18,19] for details).
Second, the symmetry corresponding to the noncommutative geometry
enforces the mass parameter to become noncommutative with {\it both}
the space and time variables (see, e.g., (3.12) below) which
is {\it different} from our earlier result in [22]. We also
study, in our present paper, 
 the deformation of the Poincar{\' e} algebra and the algebra between
the spacetime variables and the angular momenta due to the NC in spacetime
structure brought in by the non-standard gauge type transformations (and
the NC of the mass parameter with both
the space and time variables). We provide
a logical basis for the existence of the unique symmetry
transformation (see, e.g., (3.10)) for the model in the language of the local,
continuous and nilpotent
Becchi-Rouet-Stora-Tyutin (BRST) symmetry transformations and the
corresponding cohomology.

Our present study is essential primarily on three counts. First and foremost,
it is a very nice and straightforward generalization of the idea proposed
in [22] (for the  discussion of the
toy model of a reparametrization invariant non-relativistic
free particle) to the physically interesting reparametrization invariant
model of a free massive relativistic particle. Second, the 
specific type of the time-space
NC (i.e. $\{X_0, X_i\}_{(PB)} \neq 0, \{X_i, X_j \}_{(PB)} = 0$
because $\theta_{0i} \neq 0, \theta_{ij} = 0$) emerges naturally in
our present endeavour (cf. (3.1) below)
from the symmetry considerations. This type
of NC is not chosen {\it ab initio} in our 
present discussion as is the case in [15-17]. Finally,
the NC and commutativity in spacetime structure
are shown to be the distinct and different aspects of a 
unique set of symmetry transformations (cf. (3.10) below) for the
dynamical variables of the Lagrangian. The above unique symmetry 
transformations are derived from
the non-standard gauge-type symmetry transformations (corresponding to
a noncommutative geometry) and it is almost impossible to guess them
from the standard gauge symmetry transformations (corresponding to
a commutative geometry). It appears to be an interesting coincidence
that these unique symmetry transformations happen to be a
particular case of the standard gauge transformations.

The outline of our present paper is as follows. In Section 2, we recapitulate
the bare essentials of the continuous symmetry
transformations respected by the Lagrangians for the free massive
relativistic particle. Section 3 is devoted to the discussion
of the non-standard gauge-type symmetry transformations for the spacetime
variables which lead to the presence of a NC in the spacetime structure.
We discuss the deformation of the gauge algebra between the angular
momentum generator and spacetime variables in Section 4. 
Minor modification of the Poincar{\'e} algebra is also briefly sketched
in this section. We provide a cohomological origin, in Section 5,
for the NC of the spacetime structure
in the language of BRST symmetry transformations. Finally, in Section 6,
we make some concluding remarks and point out a few future directions
for further investigations.\\

\noindent
{\bf 2 Preliminary: Standard Symmetries and Commutativity}\\

\noindent
We begin with a set of three different looking
(but equivalent) Lagrangians for the free massive relativistic
particle, moving on a trajectory that is embedded in a $D$-dimensional 
flat Minkowskian target space 
\footnote{The flat Minkowskian $D$-dimensional target spacetime manifold, in
our adopted convention and notations,
is characterized by the metric $\eta_{\mu\nu} =$ diag
$(+ 1, -1, -1,.....)$ so that the scalar product $(A \cdot B)$ between two
non-null vectors $A^\mu$ and $B^\mu$ is defined as
$(A \cdot B) = \eta_{\mu\nu} A^\mu B^\nu \equiv \eta^{\mu\nu} A_\mu B_\nu
= A_0 B_0 - A_i B_i \equiv A_\mu B^\mu$. Here the Greek 
indices $\mu, \nu, \lambda.....
= 0, 1, 2, ........D-1$ correspond to the time and space directions
on the manifold and the Latin indices $i, j, k..... = 1, 2, .....D-1$
stand for the space directions only.}.
These Lagrangians are (see, e.g., [23,24]):
$$
\begin{array}{lcl}
L_{0} = m \; (\dot x^2)^{1/2}, \quad
L_{f} = p_\mu \dot x^\mu - {\displaystyle \frac{1}{2}\;e\; (p^2 - m^2), \quad
L_{s} = \frac{1}{2}\; \frac{\dot x^2}{e} + \frac{1}{2}\; e \; m^2.}
\end{array} \eqno(2.1)
$$
Some of the key common features of the above Lagrangians are 
(i) the force free ($\dot p_\mu = 0$) motion, (ii) the mass-shell condition
($p^2 - m^2 = 0$), and (iii) the reparametrization invariance under
$\tau \rightarrow \tau^\prime = f (\tau)$ where $\tau$ is the parameter
that characterizes the trajectory (i.e. the world-line) of the 
relativistic particle in the target space
and $f (\tau)$ is any arbitrary 
well-defined function of $\tau$. Except for the mass (i.e. the analogue
of the cosmological) parameter $m$, the canonically conjugate
target space phase variables (i.e. $x^\mu (\tau),\; p_\mu (\tau)$)
and the einbein field $e (\tau)$ are  functions of the monotonically
increasing parameter $\tau$ and $\dot x^\mu = (d x^\mu/ d\tau)$, etc. It is
clear from the Lagrangian $L_0$ (with the square root) and the second-order
Lagrangian $L_s$ (with $e$ in the denominator) that the following expressions 
$$
\begin{array}{lcl}
p_\mu = {\displaystyle \frac{m \dot x_\mu}{(\dot x^2)^{1/2}}} \;\equiv\;
 {\displaystyle \frac{m \dot x_\mu}{[\dot x_0^2 - \dot x_i^2]^{1/2}}},
\qquad 
e = {\displaystyle \frac{(\dot x^2)^{1/2}}{m}} \;\equiv\;
{\displaystyle \frac{[\dot x_0^2 -  \dot x_i^2]^{1/2}}{m}}, 
\end{array} \eqno(2.2)
$$
are correct. With the above inputs,  it is evident that the 
first-order Lagrangian $L_f$, re-expressed in the following long-hand form
$$
\begin{array}{lcl}
L_{f} = p_0 \dot x_0 - p_i \dot x_i - 
\frac{1}{2}\;e\; (p_0^2 - p_i^2 - m^2), 
\end{array} \eqno(2.3)
$$
is equivalent to the other Lagrangians $L_0$ and $L_s$. Even though the
above Lagrangians are equivalent and mutually consistent with one-another, 
we shall be concentrating only
on the first-order Lagrangian (2.3) for the rest of our discussions
\footnote{ The first-order Lagrangian $L_f$
is simpler in the sense that (i) it is
without any square root, and (ii) it is devoid of any variables (and their 
derivative(s) w.r.t. $\tau$) in the denominator. In addition, it is endowed
with the largest number of dynamical variables 
($x_0, x_i, p_0, p_i, e$). The latter feature of $L_f$ allows  more freedom
for theoretical discussions (connected with it) than the other
two Lagrangians $L_0$ and $L_s$.}. It can be checked that the following
symmetry transformations ($\delta_r$)
$$
\begin{array}{lcl}
\delta_r x_0 = \epsilon \dot x_0, \quad \delta_r x_i = \epsilon \dot x_i, 
\quad \delta_r p_0 = \epsilon \dot p_0, \quad 
\delta_r p_i = \epsilon \dot p_i, \quad \delta_r e = 
{\displaystyle \frac{d}{d\tau}}\; [\epsilon e],
\end{array} \eqno(2.4)
$$
generated due to the infinitesimal reparametrization transformation
$\tau \to \tau^\prime = \tau - \epsilon (\tau)$, leave the Lagrangian
(2.3) quasi-invariant because $\delta_r L_f = (d/ d\tau) [\epsilon L_f]$.
Here $\epsilon (\tau)$ is an infinitesimal local parameter for the
reparametrization transformation and $\delta_r \phi (\tau)
= \phi^\prime (\tau) - \phi (\tau)$ for the generic field
$\phi  = x_0, x_i, p_0, p_i, e$. Furthermore, it is evident that the
first-order Lagrangian (2.3) is endowed with a couple of first-class
constraints ($\Pi_e \approx 0, p_0^2 - p_i^2 - m^2 \approx 0$)
in the language of Dirac's prescription for the 
classification of constraints. Here $\Pi_e$ is the conjugate
momentum corresponding to the einbein field $e(\tau)$.
These constraints generate
\footnote{ The generator $G$ for the gauge transformations (2.5)
can be written in terms of the first-class constraints as:
$G = \dot \xi \Pi_e + (\xi/2) (p_0^2 - p_i^2 - m^2)$. The gauge
transformations $\delta_g$ for the generic field $\phi$ can be written as:
$\delta_g \phi = \{ \phi, G \}_{(PB)}$ where the canonical Poisson
brackets $\{ x_0, p_0 \}_{(PB)} = 1, 
\{x_i, p_j \}_{(PB)} = \delta_{ij}, \{ e, \Pi_e \}_{(PB)} = 1$ 
(taking into account the  rest of the brackets to be zero) have to be 
exploited for the explicit derivations.} 
the following gauge symmetry ($\delta_g$) transformations 
(with the infinitesimal parameter $\xi$)
$$
\begin{array}{lcl}
\delta_g x_0 = \xi p_0, \quad \delta_g x_i = \xi p_i, \quad
\delta_g p_0 = 0, \quad \delta_g p_i = 0, \quad \delta_g e = \dot \xi,
\end{array} \eqno(2.5)
$$
under which the Lagrangian $L_f$ remains quasi-invariant because
$\delta_g L_f = (d/d\tau) [(\xi/2) (p_0^2 - p_i^2 + m^2)]$. It is clear
that the gauge symmetry transformations (2.5) and the infinitesimal
reparametrization transformations (2.4) are equivalent for (i)
the identification $\xi = \epsilon e$, and (ii) the validity
of the equations of motion $\dot x_0 = e p_0, \dot x_i = e p_i, 
\dot p_0 = 0, \dot p_i = 0, p_0^2 - p_i^2 - m^2 = 0$.

At this juncture, it is worthwhile to lay emphasis on the 
crucial fact that the forms
of the non-trivial canonical {\it commutative}
 brackets $\{x_0, x_i \}_{(PB)} = 0,
\{ x_i, x_j \}_{(PB)} = 0$ (also 
$ \{ p_0, p_i \}_{(PB)} = 0, \{p_i, p_j \}_{(PB)} = 0$) in the untransformed
frames (corresponding to the {\it commutative} 
geometry of the spacetime) remain 
intact in the gauge transformed frames 
$$
\begin{array}{lcl}
&&x_0 \rightarrow X_0 = x_0 + \xi p_0, \qquad p_0 \to P_0 = p_0, \nonumber\\
&&x_i \rightarrow X_i = x_i + \xi p_i, \qquad p_i \to P_i = p_i,
\end{array} \eqno(2.6)
$$
as can be checked clearly
by the explicit computations $\{ X_0, X_i \}_{(PB)} = 0,
\{ X_i, X_j \}_{(PB)} = 0$ (and $\{ P_0, P_i \}_{(PB)} = 0, 
\{P_i, P_j \}_{(PB)} = 0$). This demonstrates that the gauge transformations
(2.6) generate {\it no} 
NC in the spacetime structure. This fact is reflected
in the invariance of the Poincar{\'e} algebra constructed by the
momenta $p_\mu$ and the boost $M_{0i} = x_0 p_i - x_i p_0$ and the
rotation $M_{ij} = x_i p_j - x_j p_i$ generators as given below:
$$
\begin{array}{lcl}
&&\Bigl \{ p_\mu, p_\nu \Bigr \}_{(PB)} = 0, \qquad
\Bigl \{ M_{\mu\nu}, p_\lambda \Bigr \}_{(PB)} = \delta_{\mu\lambda}
p_\nu - \delta_{\nu\lambda} p_\mu, \nonumber\\
&&
\Bigl \{ M_{\mu\nu}, M_{\lambda\zeta}  \Bigr \}_{(PB)} = \delta_{\mu\lambda}
M_{\nu\zeta} + \delta_{\nu\zeta} M_{\mu\lambda} - \delta_{\mu\zeta}
M_{\nu\lambda} - \delta_{\nu\lambda} M_{\mu\zeta},
\end{array} \eqno(2.7)
$$
where the antisymmetric generator $M_{\mu\nu} = x_\mu p_\nu - x_\nu p_\mu$
encompasses the boost $(M_{0i})$ as well as the rotation 
$(M_{ij})$ generators. The above invariance exists
 because of the gauge-invariance 
(i.e. $p_\mu^\prime \equiv P_\mu = p_\mu, M_{\mu\nu}^\prime \equiv
X_\mu P_\nu - X_\mu P_\nu = M_{\mu\nu} $) of the
generators $p_\mu$ and $M_{\mu\nu}$ 
which can be checked explicitly by using (2.6).
The above algebra should be contrasted against the Poisson brackets between
the spacetime variables $x_\mu$ (which are {\it not} gauge-invariant)
with the gauge-invariant Poincar{\'e} generators $p_\mu$ and $M_{\mu\nu}$.
For instance, it can be checked, using (2.6),
that $\{x_\mu, p_\nu\}_{(PB)} = \delta_{\mu\nu}
\rightarrow \{X_\mu, P_\nu \}_{(PB)} = 
\{x_\mu, p_\nu \}_{(PB)} = \delta_{\mu\nu}$ showing that the
algebra between gauge non-invariant spacetime variable $x_\mu$ and the 
gauge-invariant
momentum generator $p_\mu$ remain {\it invariant}. However, the algebra
between the gauge-invariant angular momenta $M_{\mu\nu}$
(i.e. $M_{\mu\nu} = x_\mu p_\nu - x_\nu p_\mu
 \to M_{\mu\nu}^\prime
= X_\mu P_\nu - X_\nu P_\mu = M_{\mu\nu}$) and the
gauge non-invariant spacetime variables $x_\mu$
(i.e. $x_\mu \to X_\mu = x_\mu + \xi \;p_\mu$):
$$
\begin{array}{lcl}
\Bigl \{M_{\mu\nu}, x_\lambda \}_{(PB)} = \delta_{\mu\lambda} x_\nu
- \delta_{\nu\lambda} x_\mu \;\to\;
\Bigl \{M_{\mu\nu}, X_\lambda \}_{(PB)} = \delta_{\mu\lambda} X_\nu
- \delta_{\nu\lambda} X_\mu,
\end{array} \eqno(2.8)
$$
remains {\it form-invariant} where $X_\mu$ is the gauge-transformed variable
defined in (2.6).

Now we shall dwell a bit on the explicit derivation
of the infinitesimal gauge symmetry transformations
(2.5) by requiring the mutual consistency among (i) the equations of motion
derived from the set of equivalent Lagrangians (2.1), (ii) the definition of
the canonical momenta $p_\mu$ from these Lagrangians, and (iii) the basic gauge
symmetry transformations (i.e. $\delta_g x_0 = \xi p_0, 
\delta_g x_i = \xi p_i$) for the space and time variables $x_i$ and $x_0$. For 
instance, it is clear from the equation (2.2) that $\delta_g e = (1/m)
\delta_g \{ [\dot x_0^2 - \dot x_i^2]^{(1/2)} \}$. Exploiting the basic gauge
transformations for the spacetime variables ($\delta_g x_\mu = \xi p_\mu$)
and using the equations of motion $\dot p_\mu = 0$ as well as the
definition of $p_\mu$ from (2.2), it can be checked
that  $\delta_g e = \dot \xi$. In a 
similar fashion, using the gauge transformations on the equations of motion
$\dot x_\mu = e p_\mu$
(i.e. $\delta_g \dot x_\mu = \delta_g\; [e p_\mu]$), it is obvious that
$\delta_g p_0 = 0, \delta_g p_i = 0$ if we exploit the equation of
motion $\dot p_\mu = 0$ and earlier derived gauge transformation 
(i.e. $ \delta_g e = \dot \xi$) for the einbein field $e$. Alternatively, the
equation of motion $\dot x_\mu = e p_\mu$ implies that
$p_\mu = (\dot x_\mu/ e)$. Exploiting the earlier derived
gauge transformation 
$\delta_g e = \dot \xi$ and the basic gauge transformations
$\delta_g x_\mu = \xi p_\mu$, we obtain
$$
\begin{array}{lcl}
\delta_g p_\mu = {\displaystyle \frac{\dot \xi}{e} \;
\Bigl (p_\mu - \frac{\dot x_\mu}{e} \Bigr ) + \frac{\xi}{e}\; \dot p_\mu,}
\end{array} \eqno(2.9)
$$
which ultimately implies: $\delta_g p_\mu = 0$ if we use the
equations of motion $\dot p_\mu = 0, \dot x_\mu = e p_\mu$. 
So far, the above gauge transformations have not been 
applied to the equations of motion 
$\dot p_\mu = 0, p_0^2 - p_i^2 - m^2 = 0$
derived from (2.3). In fact, the
explicit gauge transformations (i.e. $\delta_g x_0 = \xi p_0, \delta_g x_i
=  \xi p_i, \delta_g p_0 = 0, \delta_g p_i = 0, \delta_g e = \dot \xi$) are
consistent with the equations of motions (i.e. $\dot p_0 = 0, \dot p_i = 0$
and $p_0^2 - p_i^2 - m^2 = 0$) because $\delta_g \dot p_\mu 
\equiv (d/d\tau) [\delta_g p_\mu] = 0$ and $\delta_g [p_0^2 - p_i^2 - m^2 = 0]
\Rightarrow p_0 \delta_g p_0 
- p_i \delta_g p_i = 0$ are trivially satisfied. This technique
of the derivation of the gauge  transformations for the rest of
the fields, from a {\it given} symmetry transformation for the basic spacetime
variables, will be exploited in the next section.\\

\noindent
{\bf 3 Noncommutativity and Non-Standard Symmetries}\\

\noindent
Similar to the standard gauge transformations (2.6), let us pay
attention to the 
following non-standard spacetime transformations
(with an infinitesimal parameter $\zeta(\tau)$)
$$
\begin{array}{lcl}
&&x_0 \to X_0 = x_0 + \zeta \;\theta_{0i}\; p_i \;\Rightarrow \;
\tilde \delta_g x_0 = \zeta\; \theta_{0i}\; p_i, \nonumber\\
&& x_i \to X_i = x_i + \zeta\; \theta_{i0} \;p_0 \;\Rightarrow\;
\tilde \delta_g x_i = \zeta \;\theta_{i0} \;p_0, 
\end{array} \eqno(3.1)
$$
which imply a time-space NC because the nontrivial Poisson bracket for the
transformed spacetime variables turns out to be non-zero
(i.e. $ \{X_0, X_i\}_{(PB)} = - 2 \zeta \theta_{0i} $). In the above
derivation, we have (i) treated the antisymmetric
(i.e. $\theta_{0i} = - \theta_{i0}$) parameter $\theta_{0i}$ to be 
a constant (i.e. independent of the parameter
$\tau$ as well as the phase space variables), and
(ii) exploited the canonical brackets $\{ x_\mu, x_\nu \}_{(PB)} = 0,
\{p_\mu, p_\nu \}_{(PB)} = 0, \{x_\mu, p_\nu \}_{(PB)} = \delta_{\mu\nu}$.
It is elementary to check that $\{X_0, X_0\}_{(PB)} = \{X_i, X_j\}_{(PB)} = 0$.
One can treat the above NC to be a special case of the general NC
defined through $\{X_\mu (\tau), X_\nu (\tau) \}_{(PB)} 
= \Theta_{\mu\nu} (\tau)$ on the
spacetime target manifold where $\Theta_{0i} (\tau) = - 2 \zeta (\tau) 
\theta_{0i}, \Theta_{ij} (\tau) = - 2 \zeta (\tau) \theta_{ij} = 0$
\footnote{We shall furnish, later in the present section,
a proof of $\theta_{ij} = 0 (\Rightarrow \Theta_{ij} = 0)$ by taking into
account the consistency between the symmetry properties and the equations
motion for the model under consideration.}. In fact, 
such a kind of NC has been discussed extensively in [15-17]. The special type
of transformations (3.1) have been taken into account primarily
for a couple of reasons. First, they lead to the time-space NC
(i.e. $\theta_{0i} \neq 0, \theta_{ij} = 0$)
in the transformed spacetime manifold which has been used, in detail,
 for the development of a unitary quantum mechanics [15]. Second, 
they are relevant
in the context of BRST symmetry transformations and corresponding
cohomology (see, e.g., Section 5 below for a detailed discussion).

Taking into account the basic transformations (3.1) on the spacetime
variables and demanding their consistency with some of the equations of motion,
derived from the set of Lagrangians (2.1), we obtain (using the trick 
discussed earlier in connection with the derivation of the standard gauge
transformations) the following
non-standard transformations for the rest of the dynamical variables 
of the Lagrangian $L_f$ of (2.3), namely;
$$
\begin{array}{lcl}
\tilde \delta_g e &=& {\displaystyle 
\frac{ 2\; \dot \zeta\; \theta_{0i}\; p_0\; p_i}{m^2} \equiv
\frac{ 2\; \dot \zeta\; \theta_{0i}\; \dot x_0\; \dot x_i}{\dot x^2}},
\nonumber\\
\tilde \delta_g p_0 &=& {\displaystyle 
\frac{ 2 \;\dot \zeta \; \theta_{0i} \;p_i}{e} \; \Bigl [\; \frac{1}{2} - 
\frac{p_0^2}{m^2}\; \Bigr ]}, \nonumber\\
\tilde \delta_g p_i &=& - {\displaystyle 
\frac{ 2 \;\theta_{0j}\; p_0\; \dot \zeta}{e} \; \Bigl [\; 
\frac{\delta_{ij}}{2} +
\frac{p_i p_j}{m^2} \;\Bigr ]}. 
 \end{array} \eqno(3.2)
$$
It should be noted that, so far, the above transformations have {\it not} been
applied to (and contrasted against) the sanctity of the equations of motion
$\dot p_0 = 0, \dot p_i = 0, p_0^2 - p_i^2 - m^2 = 0$. It can be 
checked that the above
non-standard transformations are consistent with $p_0^2 - p_i^2 = m^2$ because
the relation $\tilde \delta_g p_0 = (1/p_0) [p_i \tilde \delta_g p_i]$ is 
satisfied without any restriction on any parameters. However, the story is
totally different as far as the
consistency with $\tilde \delta_g \dot p_\mu = 0$
is concerned. It can be seen that, on the on-shell 
$\dot p_0 = 0, \dot p_i = 0$, we have
$$
\begin{array}{lcl}
{\displaystyle \frac{d}{d \tau} (\tilde \delta_g p_0) = 0
\Rightarrow \frac{2}{e^2} \; \Bigl [ \ddot \zeta e 
- \dot \zeta \dot e \Bigr ]\;
\Bigl [\theta_{0i}\; p_i \Bigl (\frac{1}{2} - \frac{p_0^2}{m^2} \Bigr )
\Bigr ]} = 0,
\end{array} \eqno(3.3)
$$
which can be satisfied (for $ e \neq 0$) by any (or all) 
of the following conditions
$$
\begin{array}{lcl}
(i)\;\; \theta_{0i} p_i = 0, \qquad (ii) \;\;
\ddot \zeta e - \dot \zeta \dot e = 0,
\qquad (iii) \;\;p_0^2 = m^2/2.
\end{array} \eqno(3.4)
$$
It is evident that $\theta_{0i} p_i = 0$ is {\it not} an interesting solution
because, in this case, there is no transformation (i.e. 
$\tilde \delta_g x_0 = \theta_{0i} p_i = 0$) for
the time variable $x_0$. As a result, there will be no  
time-space NC in the theory. Before we shall focus on the solutions
(ii) and (iii) of (3.4), let us find out the restrictions from the
following (i.e. $\tilde \delta_g [\dot p_i] = 0$) consistency condition
$$
\begin{array}{lcl}
{\displaystyle \frac{d}{d \tau} (\tilde \delta_g p_i) = 0
\Rightarrow - \frac{2}{e^2} \; \Bigl [ \ddot \zeta e 
- \dot \zeta \dot e \Bigr ]\;
\Bigl [\theta_{0j}\; p_0 \Bigl (\frac{\delta_{ij}}{2} +
 \frac{p_i p_j}{m^2} \Bigr )
\Bigr ]} = 0.
\end{array} \eqno(3.5)
$$
It is clear that, for $e \neq 0$, the above restrictions are satisfied by the 
solution (ii) of (3.4) and, in addition, there is yet another restriction
as given below
$$
\begin{array}{lcl}
{\displaystyle \frac{\theta_{0i} p_0}{2} + p_i\; 
\frac{\theta_{0j} p_0 p_j} {m^2}} = 0.
\end{array} \eqno(3.6)
$$
It is evident now that the conditions to be satisfied are (3.6) as well as
(ii) and (iii) of (3.4) so that the restrictions (3.3) and (3.5) could be
satisfied {\it together}.

Let us now embark on the explicit solutions to (ii) of equation (3.4)
which is a common condition for the solution to (3.3) and (3.5). As a side
remark, it is worthwhile to mention that such a condition has also
been obtained in our earlier work [22] on reparametrization invariant toy
model of a  non-relativistic particle. We recapitulate here some of the
key points of our arguments in [22]. It is evident, from the
condition (ii) of (3.4), that $(\ddot \zeta/\dot \zeta) = (\dot e/ e) = R$
where the ratio $R$ is some well-defined function of $\tau$ (or a constant).
Let us choose $R$ to be a constant: $R = \pm K$ where $K$ is independent
of $\tau$. The solutions that emerge for $e$ and $\zeta$ are:
$e (\tau) = e(0) e^{\pm K \tau}, \zeta (\tau) = \zeta (0) e^{\pm K \tau}$.
Unfortunately, these values, when substituted in (3.2), lead to no
interesting symmetry transformations
for the first-order Lagrangian $L_f$. Further,
a simpler non-trivial choice $R = \pm \tau$ leads to the solutions for
$e$ and $\zeta$ as: $e(\tau) = e(0) e^{\pm (\tau^2/2)}$ and a series
solution $\zeta (\tau) = \tau \pm \sum_{n = 1}^{\infty}\; [(2n - 1)!!]/
[ (2n + 1)!]\; \tau^{2n + 1}$. The substitution of these values in
(3.2), once again, does not lead to any worthwhile symmetry property
of $L_f$. Thus, we are led to conclude that, even though, the restriction
(ii) of (3.4), is a common solution to (3.3) and (3.5), its thorough
discussion does not lead to any interesting symmetry property of $L_f$.
To be precise, we do not know, at the moment, any worthwhile solution
to the condition (ii) of (3.4) that leads to some interesting
symmetry property of $L_f$. At present, it is an open problem to find out
an interesting solution to (ii) of equation (3.4) (which happens to be
one of the solutions to (3.5), too.)

We shall now focus on the left over conditions $p_0^2 = (m^2/2)$
and (3.6). The former one trivially 
implies that $p_0 = \pm (m/\surd 2)$. However, to obtain the solution to
the latter one, our previous knowledge of our earlier work [22] on
the reparametrization invariant free non-relativistic particle
turns out to be quite handy. It is clear that if we make the following 
choice in (3.6)
$$
\begin{array}{lcl}
\theta_{0i}\; p_0\; p_i = {\displaystyle \frac{m^2}{2}}, 
\end{array} \eqno(3.7)
$$
a couple of interesting consequences emerge very naturally. First, we notice
that the transformation on the einbein field becomes the gauge transformation
(i.e. $\tilde \delta_g \to \delta_{g}$) of (2.5) with the identification
$\zeta (\tau)  = \xi (\tau)$); namely
$$
\begin{array}{lcl}
\tilde \delta_g \;e = {\displaystyle 
\frac{2\; \dot \zeta \;\theta_{0i}\; p_0 \;p_i}{m^2}} \;\;\to \;\;
\delta_{g} \; e
= \dot \zeta \equiv \dot \xi.
\end{array} \eqno(3.8)
$$
Second, the relation in (3.6) with the above choice, ultimately, 
leads to (i) a connection between $p_0$ and $p_i$ through
noncommutative parameter $\theta_{0i}$, and (ii) the noncommutative 
parameters $\theta_{0i}$ are restricted. These observations
are captured in the following relations:
$$
\begin{array}{lcl}
p_i = \theta_{i0} p_0, \qquad \theta_{0i} \theta_{0j} = 
\theta_{i0} \theta_{j0} = - \delta_{ij},
\qquad p_0 = \theta_{0i} p_i, \qquad p_i p_j = - 
{\displaystyle \frac{m^2}{2} \delta_{ij}}.
\end{array} \eqno(3.9)
$$
It is now obvious that the non-standard transformations (3.1) and (3.2)
reduce {\it exactly} to a particular case of
the standard gauge transformations (2.5). However,
there is a catch. The zeroth component of the momentum has now been
fixed (i.e. $ p_0 = + (m /\surd 2)$)
\footnote{The choice $p_0 = - (m/\surd 2)$, 
implying $p_i = - (m/\surd 2) \theta_{i0}$, does also lead to a symmetry
transformation. However, the negative energy ($p_0 = - (m/\surd 2)$ solution
is not a physically interesting choice for a ``particle''.}. 
As a consequence, the space component of
momentum is also fixed as: $p_i = (m /\surd 2)\; \theta_{i0}$. Thus,
an interesting  limiting case (i.e. $\delta_{g} \to \delta_{g}^{(sp)}$)
of the gauge transformations (2.5) emerges, 
from the non-standard transformations (3.1) and (3.2) 
(i.e. $\tilde \delta_g \to \delta_g^{(sp)}$ with the identification
($ \xi (\tau) = \zeta (\tau)$), namely;
$$
\begin{array}{lcl}
\delta^{(sp)}_g x_0 = {\displaystyle \frac{m}{\surd 2} \xi, \quad 
\delta^{(sp)}_g x_i = \frac{m}{\surd 2}} \xi \theta_{i0}, \quad
\delta^{(sp)}_g p_0 = 0, \quad \delta^{(sp)}_g p_i = 0, \quad 
\delta^{(sp)}_g e = \dot \xi.
\end{array} \eqno(3.10)
$$
At this juncture, there are a few noteworthy points that have to be
mentioned. First of all, we note that $\delta_{g}^{(sp)}$ emerges
from (2.5) as well as from 
(3.1) and (3.2). Thus, there are a couple of ways to 
interpret the symmetry transformations (3.10). Second, the values
for $p_0 = (m/\surd 2)$ and $p_i = (m/\surd 2)\;\theta_{i0}$, obtained
from the consistency requirements, satisfy all the equations of motion
(i.e. $\dot p_0 = 0, \dot p_i = 0, p_0^2 - p_i^2 = m^2$) and the choice
(3.7) with inputs from (3.9). Third, with the above values of $p_0$
and $p_i$, the first-order Lagrangian $L_f$ in (2.3) reduces effectively to
the following simpler looking  form
$$
\begin{array}{lcl}
L_{f1}^{(eff)} = {\displaystyle \frac{m} {\surd 2} \;\dot x_0
- \frac{m}{\surd 2}\; \dot x_i\; \theta_{i0} }.  
\end{array} \eqno(3.11)
$$
The above Lagrangian 
 remains quasi-invariant (i.e. $\delta_{g}^{(sp)} L_{f1}^{(eff)}
= (d/d\tau)\; [ m^2 \xi ]$) under the symmetry transformations (3.10).
There is yet another way to obtain a simpler form of (2.3) (with the inputs
$p_0 = (m/\surd 2)$ and $p_i = (m/\surd 2) \theta_{i0}$) because
now $\dot x_0 = e p_0 \equiv (m/\surd 2)\; e$ and $\dot x_i = (m/\surd 2)\;
e\; \theta_{i0}$ imply that the above effective Lagrangian (3.11) can be
re-expressed as: $L_{f2}^{(eff)} = m^2 \;e$. It is evident that,
under the transformations (3.10), the effective Lagrangian $L_{f2}^{(eff)}$
also remains quasi-invariant because 
$\delta_g^{(sp)} L_{f2}^{(eff)} = (d/d\tau)\; [m^2 \xi]$.
It is very interesting to note that our starting Lagrangian
$L_0 = m (\dot x^2)^{(1/2)}$ also transforms to
$\delta_g  L_0 = (d/d\tau)\; [m^2 \xi]$ under the usual gauge transformation
$\delta_g x_\mu = \xi \; [m \dot x_\mu/ (\dot x^2)^{(1/2)}]$.

Now we shall dwell a bit on the NC and commutativity properties of the present
model of the massive relativistic particle. It has been shown 
in [18,19], through Dirac
bracket formalism, that the NC and commutativity for this model owe their
origin to the different choices of the gauge-condition. These 
gauge-conditions, in turn, are connected to each-other by some kind
of a gauge 
transformation. Thus, the NC and commutativity of this model, 
in some sense, are equivalent (because one can gauge away the NC by
the redefinition of the variables [18]). This result can be explained
in the language of symmetry properties of the Lagrangian $L_f$.
It is clear from the gauge transformations (2.6) that there is no
spacetime NC in the (un-)transformed frames. Thus, equation (3.10),
which can be derived from the gauge transformation (2.5) with
the substitution $p_0 = (m/\surd 2)$ and $p_i = (m/\surd 2)\;\theta_{i0}$,
implies {\it no} NC at all. However, if we focus on (3.10) from the
point of view of (3.1) and (3.2) (with conditions listed in (3.9)), we find 
that there exists a NC in the transformed frames of spacetime which entails
upon the mass parameter $m$ to become {\it noncommutative} in nature. Let us
recall that, the NC present
in the transformed frames (i.e. $\{ X_0, X_i \}_{(PB)}
= - 2 \zeta \theta_{0i}$) corresponding to the transformations (3.1),
basically owes its origin to the canonical brackets $\{x_0, p_0 \}_{(PB)}
= 1, \{x_i, p_j \}_{(PB)} = \delta_{ij}$. Thus, it is clear that,
with the solutions $p_0 = (m/\surd 2), p_i = (m/\surd 2) \theta_{i0},
\theta_{0i} \theta_{0j} = - \delta_{ij}$, 
these brackets reduce  to 
$$
\begin{array}{lcl}
\Bigl \{ x_0, m \Bigr \}_{(PB)} = \surd 2, \;\;\qquad\;\;
\Bigl \{ x_i, m \Bigr \}_{(PB)} = - \surd 2 \; \theta_{i0}, 
 \end{array} \eqno(3.12)
$$
demonstrating that the original NC associated with the transformations (3.1)
enforces the mass parameter to become noncommutative. It is 
interesting to emphasize that the NC of the mass parameter $m$ with the
time (i.e. $x_0$) and the space (i.e. $x_i$) variables is primarily due
to the connections (i.e. $p_i = \theta_{i0} p_0, p_0 = \theta_{0i} \; p_i$)
between the momentum  components $p_0$ and $p_i$ through the 
antisymmetric parameter $\theta_{0i}$. Thus,
the above arguments establish that the NC and commutativity of the model
under consideration are equivalent and are different facets of a specific
gauge symmetry transformation (cf. (3.10)).
The NC of mass parameter is {\it not} a new observation. In the realm
of quantum groups, the NC nature of mass parameter has been shown in
the context of free motion of the (non-)relativistic particles on a 
quantum-line, embedded in a cotangent manifold [25-28].

Before we wrap up this section, we would like to shed some light on
the condition  $\theta_{ij} = 0$ from the requirement of the consistency
between the symmetry properties and the equations of motion for the 
model under consideration. Towards this goal in mind, let us begin with
a more general non-standard spacetime transformation $\tilde \delta_g^{(1)}$
$$
\begin{array}{lcl}
\tilde \delta_g^{(1)}  x_0 = \zeta_1 \;\theta_{0i}\; p_i, \qquad
\tilde \delta_g^{(1)}  x_i = \zeta_1\; (\theta_{i0}\; p_0
+ \theta_{ij}\; p_j),
\end{array} \eqno(3.13)
$$
where 
$\zeta_1 (\tau)$ is an infinitesimal transformation parameter and 
$\theta_{ij} \neq 0$  so that 
$\{X_i, X_j \}_{(PB)} = - 2 \zeta_1 \theta_{ij}$.
It is elementary to check that, in the limit
$\theta_{ij} = 0$, we get back the non-standard transformations (3.1).
Given the above spacetime transformations, we have to derive the 
transformations for the rest of the variables of the Lagrangian $L_f$
(cf. (2.3)). We adopt here the same technique as we have exploited for
the derivation of (2.5) and (3.2). These explicit transformations, 
which are the analogue of (3.2), are as follows:
$$
\begin{array}{lcl}
\tilde \delta^{(1)}_g e &=& {\displaystyle 
\frac{ 2\; \dot \zeta_1\; \theta_{0i}\; p_0\; p_i}{m^2} \equiv
\frac{ 2\; \dot \zeta_1\; \theta_{0i}\; \dot x_0\; \dot x_i}{\dot x^2}},
\nonumber\\
\tilde \delta^{(1)}_g p_0 &=& {\displaystyle 
\frac{ 2 \;\dot \zeta_1 \; \theta_{0i} \;p_i}{e} \; \Bigl [\; \frac{1}{2} - 
\frac{p_0^2}{m^2}\; \Bigr ]}, \nonumber\\
\tilde \delta^{(1)}_g p_i &=& - {\displaystyle 
\frac{ 2 \;\theta_{0j}\; p_0\; \dot \zeta_1}{e} \; \Bigl [\; 
\frac{\delta_{ij}}{2} +
\frac{p_i p_j}{m^2} \;\Bigr ] + \frac{ \dot \zeta_1\; \theta_{ij}\; p_j}
{e}}. 
 \end{array} \eqno(3.14)
$$
The above transformations are, once again, consistent with the equation
of motion $p_0^2 - p_i^2 - m^2 = 0$ because it can be checked that
$\tilde \delta^{(1)}_g p_0 = (1/ p_0)\; [p_i \tilde \delta^{(1)}_g p_i]$
is satisfied without any restriction on any of the parameters or fields
of the theory. However, the consistency with 
$\tilde \delta^{(1)}_g \dot p_0
= (d/ d\tau) [\tilde \delta^{(1)}_g p_0] = 0$ and
$\tilde \delta^{(1)}_g \dot p_i
= (d/ d\tau) [\tilde \delta^{(1)}_g p_i] = 0$ leads to the following
additional restriction besides the restrictions given in (3.3) and (3.4),
namely;
$$
\begin{array}{lcl}
{\displaystyle \frac{d}{d \tau} (\tilde \delta^{(1)}_g p_i) = 0
\Rightarrow - \frac{2}{e^2} \; \Bigl [ \ddot \zeta_1 e 
- \dot \zeta_1 \dot e \Bigr ]\;
\Bigl [\theta_{0j}\; p_0 \Bigl (\frac{\delta_{ij}}{2} +
 \frac{p_i p_j}{m^2} \Bigr ) - \frac{\theta_{ij} p_j}{2}
\Bigr ]} = 0,
\end{array} \eqno(3.15)
$$
which is nothing but the generalization of (3.5). Analogous discussions, as
given earlier, demonstrate that the generalized version of (3.6) is now
$$
\begin{array}{lcl}
{\displaystyle \frac{\theta_{0i} p_0}{2} + p_i\; 
\frac{\theta_{0j} p_0 p_j} {m^2} = \frac{\theta_{ij} p_j}{2}}.
\end{array} \eqno(3.16)
$$
A couple of points are clear from the validity of (3.4), even in the 
present case of our discussions, where $\theta_{ij} \neq 0$. First,
the solution (i) of (3.4) is not allowed as argued earlier and
the condition (ii) of (3.4) does not lead to any interesting symmetry
properties of $L_f$. Second, the choice $p_0 = (m/\surd 2)$ is still
an interesting solution emerging from (3.4). Our aim is to obtain a 
gauge-type transformation from (3.14), too. For this purpose, we summon
our knowledge of the previous work [22] and make the choice as given
in (3.7) so that we can obtain the usual gauge transformation
($\delta_g e = \dot \xi$) for the einbein field $e (\tau)$
with the identification $\zeta_1 (\tau) = \xi (\tau)$.  The substitution
of the values from (3.4) and (3.7), into the condition (3.16), leads
to $\theta_{ij} p_j = 0$. For the generality of the transformations,
it is proper to assume that $p_i \neq 0$. Thus, a logical conclusion is
$\theta_{ij} = 0$ which has been taken into account for the discussion
of the unitary quantum mechanics [15]. This establishes the fact that
the consistency requirement between the equations of motion and
the exact symmetry properties enforces the NC parameter in
$\{X_\mu (\tau), X_\nu (\tau)\}_{(PB)} = \Theta_{\mu\nu} (\tau)$
(with $\Theta_{\mu\nu} (\tau) = - 2 \zeta (\tau) \theta_{\mu\nu}$)
to possess only the component $\Theta_{0i} (\tau) = - 2 \zeta (\tau)
\theta_{0i}$ and the component 
$\Theta_{ij} = - 2 \zeta (\tau) \theta_{ij}$
is zero due to $\theta_{ij} = 0$.
It is now obvious that the transformations (3.1) and (3.2), with
certain restrictions emerging from the consistency conditions
between the equations of motion and symmetry properties, are the 
only allowed transformations for our present model.\\

\noindent
{\bf 4 Deformations of the Algebras}\\

\noindent
Let us now concentrate on the usual
Poincar{\'e} algebra (2.7) and an additional
algebra (2.8) between the angular momentum generator ($M_{\mu\nu}$)
and the spacetime variable ($x_\mu$).
As mentioned earlier, the angular momentum ($M_{\mu\nu}$) and
momentum ($p_\mu$) generators remain invariant under the usual
gauge transformations (2.6) (or (2.5)). In other words, these generators
and their algebra remain {\it invariant} in the (un-)transformed frames.
This statement is true even with the transformations given in (3.10)
which have been obtained from the non-standard gauge-type symmetry
transformations in (3.1) and (3.2). However, the story is quite different
with the additional algebra (2.8). To observe the impact of NC on this
algebra, let us express the algebra (2.8) in the component form
with the boost generator $M_{0i}$ and the rotation generator $M_{ij}$.
The set of these Poisson brackets, in explicit form, are
$$
\begin{array}{lcl}
&&\Bigl \{M_{0i}, x_j \}_{(PB)} = 
- \delta_{ij} x_0,\; \qquad \;
\Bigl \{M_{0i}, x_0 \}_{(PB)} = x_i, \nonumber\\
&&\Bigl \{M_{ij}, x_0 \}_{(PB)} = 0,\;\; \qquad\;\;
\Bigl \{M_{ij}, x_k \}_{(PB)} = \delta_{ik} x_j - \delta_{jk} x_i.
\end{array} \eqno(4.1)
$$
It is evident, from (2.8) itself, that the algebra (4.1) remains
{\it form-invariant} in the gauge-transformed frame where
$x_0 \to X_0, x_i \to X_i$ (cf. (2.6)) and $M_{0i} \to M_{0i},
M_{ij} \to M_{ij}$. Thus, in the transformed frame, the above algebra
will be replaced by an algebra where $x_0 \to X_0$ and $x_i \to X_i$ 
(cf. (2.6)). In the noncommutative case, corresponding to the 
transformations (3.10), the transformed spacetime variables 
$(X^{(sp)}_i, X^{(sp)}_0)$
and the gauge invariant boost ($M_{0i}$) and rotation ($M_{ij}$)
generators are
$$
\begin{array}{lcl}
&&x_0 \to X^{(sp)}_0 = x_0 + \xi {\displaystyle \frac{m}{\surd 2}}, \qquad 
x_i \to X^{(sp)}_i = x_i + \xi \theta_{i0} {\displaystyle \frac{m}{\surd 2}}, 
\nonumber\\
&& M_{0i} = {\displaystyle 
x_0 \theta_{i0}
\frac{m}{\surd 2} 
- x_i \frac{m}{\surd 2},\; \qquad\;
 M_{ij} = 
x_i \theta_{j0} \frac{m}{\surd 2} 
- x_j \theta_{i0} \frac{m}{\surd 2}},
\end{array} \eqno(4.2)
$$
where we have used $p_0 = (m/\surd 2)$ and $p_i = \theta_{i0} (m/\surd 2)$.
With the inputs from (4.2), it can be checked that the algebra (4.1)
(in the component form) changes to the following:
$$
\begin{array}{lcl}
&&\Bigl \{M_{0i}, x_j \}_{(PB)} = 
- \delta_{ij} x_0 - x_i \theta_{j0},\;\; \qquad \;\;
\Bigl \{M_{0i}, x_0 \}_{(PB)} = x_i - \theta_{i0} x_0, \nonumber\\
&&\Bigl \{M_{ij}, x_0 \}_{(PB)} = x_j \theta_{i0} - x_i \theta_{j0}, \qquad
\Bigl \{M_{ij}, x_k \}_{(PB)} = \delta_{ik} x_j - \delta_{jk} x_i.
\end{array} \eqno(4.3)
$$
There are a few comments in order now. First, it can be seen that,
in the limit $\theta_{i0} \to 0$, we do get back the algebra (4.1)
for the usual commutative spacetime. Second, we have exploited
the NC of the mass parameter through the Poisson brackets (3.12)
(and their off-shoots $\{m, x_0 \}_{(PB)} = - \surd 2, 	
\{m, x_i \}_{(PB)} = + \surd 2 \theta_{i0}$). Third, it is the NC of the
mass parameter $m$ with {\it both} 
the space $x_i$ and time $x_0$ variables (cf. (3.12)) which
is the root cause of the deformation of the algebra (4.1) to (4.3). Fourth, 
we have exploited appropriately (3.9) in the above computation of the
Poisson brackets (4.3) so that its {\it form} could 
be compared with (and contrasted against) (4.1). Fifth, it can be checked that,
in the transformed frames (with spacetime variables 
$X_i^{(sp)}$ and $ X_0^{(sp)}$),
the algebra (4.3) remains {\it form-invariant} in the sense that
one has to merely replace: $x_0 \to X_0^{(sp)}, x_i \to X_i^{(sp)}$.

Now let us pay attention to the impact of the NC on the modification of 
Poincar{\'e} algebra (2.7). It is straightforward to note that the
algebra $\{p_\mu, p_\nu \}_{(PB)} = 0$ remains intact with the choice
$p_0 = (m/\surd 2), p_i = (m/\surd 2) \theta_{i0}$. However, there is some
modification (due to the NC in spacetime)
 in the algebra between the momentum generator $p_\mu$ and
the angular momentum $M_{\mu\nu}$ when they are written in the component
form. For instance, using the expression for the boost and rotation 
generators from (4.2), it can be checked that
$$
\begin{array}{lcl}
&&\Bigl \{M_{0i}, p_j \}_{(PB)} = 
- 2 \delta_{ij} p_0,\; \;\;\;\qquad \;\;\;\;\;
\Bigl \{M_{0i}, p_0 \}_{(PB)} = 2 p_i, \nonumber\\
&&\Bigl \{M_{ij}, p_0 \}_{(PB)} 
= \theta_{i0} p_j - \theta_{j0} p_i \equiv 0, \qquad
\Bigl \{M_{ij}, p_k \}_{(PB)} = \delta_{ik} p_j - \delta_{jk} p_i.
\end{array} \eqno(4.4)
$$
It should be noted that the r.h.s. (i.e. $\theta_{i0} p_j - \theta_{j0} p_i$)
of the bracket $\{M_{ij}, p_0 \}_{(PB)}$ can be proved to be zero if
we take into account the inputs (i.e. $p_i = \theta_{i0} (m/\surd 2), 
\theta_{i0} \theta_{j0} = - \delta_{ij}$) from (3.9).
In the usual case of the commutative spacetime, 
the algebra  corresponding to (4.4), that emerges from (2.7),
is as follows:
$$
\begin{array}{lcl}
&&\Bigl \{M_{0i}, p_j \}_{(PB)} = 
-  \delta_{ij} p_0,\; \qquad \;
\Bigl \{M_{0i}, p_0 \}_{(PB)} =  p_i, \nonumber\\
&&\Bigl \{M_{ij}, p_0 \}_{(PB)} = 0,\; \qquad
\Bigl \{M_{ij}, p_k \}_{(PB)} = \delta_{ik} p_j - \delta_{jk} p_i.
\end{array} \eqno(4.5)
$$
At this juncture, a couple of remarks are in order. First, the root cause
of the difference of factor two, in the above equations (4.4)
and (4.5), is the NC
of the mass parameter $m$ with both the space $x_i$ and time $x_0$ variables.
This is why, a factor of two appears in the Poisson brackets wherever the
boost ($M_{0i}$) generators are present. This factor does not turn up in the
Poisson brackets with the rotation generator $M_{ij}$.
Second, the Poisson brackets (3.12) (and their antisymmetric versions),
together with the conditions (3.9), have been appropriately exploited
to express the structure  of (4.4) so that it could be compared with the
usual form of algebra in (4.5). The triplet of the Poisson brackets
among the boost $M_{0i}$ and rotation $M_{ij}$ generators,
that emerges from the usual Poincar{\'e} algebra (2.7), are
$$
\begin{array}{lcl}
&&
\Bigl \{ M_{ij}, M_{kl}  \Bigr \}_{(PB)} = \delta_{ik}
M_{jl} + \delta_{jl} M_{ik} - \delta_{il}
M_{jk} - \delta_{jk} M_{il}, \nonumber\\
&&
\Bigl \{ M_{ij}, M_{0k}  \Bigr \}_{(PB)} = \delta_{ik}
M_{0j} - \delta_{jk} M_{0i},\; \qquad \;
\Bigl \{ M_{0i}, M_{0j}  \Bigr \}_{(PB)} = M_{ij}.
\end{array} \eqno(4.6)
$$
Let us now compute the {\it extra} contribution 
to the above algebra when the NC of spacetime is taken into account.
For this purpose, we have to plug in the exact expressions
for the boost and rotation generators from (4.2). The explicit
computation of the Poisson brackets, using (3.12)
and its antisymmetric versions, leads to the following:
$$
\begin{array}{lcl}
&&
\Bigl \{ M_{ij}, M_{kl}  \Bigr \}_{(PB)} = \delta_{ik}
M_{jl} + \delta_{jl} M_{ik} - \delta_{il}
M_{jk} - \delta_{jk} M_{il}, \nonumber\\
&&
\Bigl \{ M_{ij}, M_{0k}  \Bigr \}_{(PB)} = 2 \;\Bigl [\;\delta_{ik}
M_{0j} - \delta_{jk} M_{0i}\; \Bigr ], \;\;\qquad \;\;
\Bigl \{ M_{0i}, M_{0j}  \Bigr \}_{(PB)} = 2 M_{ij}.
\end{array} \eqno(4.7)
$$
It is clear, once again, that the extra factor of 2 appears
in the case of the Poisson brackets where the boost generators 
$M_{0i}$ are present (and the NC of the mass parameter $m$ with
both the space and time variables is taken into account). It is
obvious that all the algebras, from (4.4) to (4.7), remain
{\it invariant} in the (un-)transformed frames. \\

\noindent
{\bf 5 Noncommutativity and (Anti-)BRST Symmetries }\\

\noindent
As discussed earlier, the first-class constraints for the Lagrangian
$L_f$ in (2.3), generate a set of
``classical'' local gauge symmetry transformations
(2.5) for all the fields. These ``classical'' gauge symmetries can be traded
with the ``quantum'' gauge symmetries which are popularly known
as the BRST symmetries. These are the  generalized ``quantum'' versions of
the local gauge symmetries and are found to be nilpotent of order two.
In the BRST formalism, the unitarity and the ``quantum'' gauge
(i.e. BRST) invariance are respected
together at any arbitrary order of perturbative computations for
a given physical process in a BRST
invariant gauge theory [29]. The (anti-)BRST invariant version of the
local gauge-invariant Lagrangian (2.3) is (see, e.g., [24])
$$
\begin{array}{lcl}
L_b = p_0 \dot x_0 - p_i \dot x_i - \frac{1}{2} \;e\; (p_0^2 -p_i^2 - m^2) 
+ b \;\dot e  + \frac{1}{2}\; b^2 - i \;\dot {\bar c} \;\dot c,
 \end{array} \eqno(5.1)
$$
where $b$ is the Nakanishi-Lautrup auxiliary field and $(\bar c)c$ are
the anticommuting (i.e. $c^2 = \bar c^2 = 0, c \bar c + \bar c c = 0$)
(anti-)ghost fields which are required in the theory to maintain
the unitarity (see, e.g., [29] for details). The above Lagrangian
$L_b$ remains quasi-invariant under the following off-shell nilpotent
($s_{(a)b}^2 = 0$) and anticommuting ($s_b s_{ab} + s_{ab} s_b = 0$)
(anti-)BRST transformations $s_{(a)b}$
\footnote{We adopt here the notations and conventions used by Weinberg [30].
In fact, in its totality, the nilpotent ($\delta_{(a)b}^2 = 0$)
(anti-)BRST transformations $\delta_{(a)b}$ are product of an anticommuting 
($ \eta c + c \eta = 0$, etc.) spacetime 
independent parameter $\eta$ and $s_{(a)b}$ with $s_{(a)b}^2 = 0$. The
(anti-)BRST prescription is to replace the local gauge parameter $\xi$  of
the gauge transformation (2.5) by $\eta$ and the (anti-)ghost fields 
$(\bar c)c$.}
$$
\begin{array}{lcl} 
&& s_b x_0 = c p_0,  \qquad s_b x_i = c p_i, \qquad
s_b p_0 = 0, \qquad s_b p_i = 0, \nonumber\\
&& s_b c = 0, \;\;\qquad\;
s_b e = \dot c, \;\;\qquad\; s_b \bar c = i b,  \;\;\qquad\; s_b b = 0,
\end{array} \eqno(5.2)
$$
$$
\begin{array}{lcl} 
&&s_{ab} x_0 = \bar c p_0, \qquad s_{ab} x_i = \bar c p_i,
\qquad s_{ab} p_0 = 0, \qquad s_{ab} p_i = 0, \nonumber\\
&&s_{ab} \bar c = 0, \qquad
s_{ab} e = \dot {\bar c}, \qquad
s_{ab} c = - i b, \qquad s_{ab} b = 0,
\end{array} \eqno(5.3)
$$
which are the ``quantum'' generalization of the 
``classical'' local gauge transformations
(2.5). The above off-shell nilpotent transformations
\footnote{The on-shell ($\ddot c = \ddot {\bar c} = 0$) nilpotent
($\tilde s_{(a)b}^2 = 0$) version of
the (anti-)BRST transformations $\tilde s_{(a)b}$:
$\tilde s_b x_0 = c p_0,  \tilde s_b x_i = c p_i, \tilde s_b p_0 = 0, 
\tilde s_b p_i = 0, \tilde s_b c = 0, \tilde s_b e = \dot c, 
\tilde s_b \bar c = -i \dot e$ and
$\tilde s_{ab} x_0 = \bar c p_0, \tilde s_{ab} x_i = \bar c p_i,
\tilde s_{ab} p_0 = 0, \tilde s_{ab} p_i = 0, \tilde s_{ab} \bar c = 0, 
\tilde s_{ab} e = \dot {\bar c}, \tilde s_{ab} c = i \dot e$ do exist
for the Lagrangian
$\tilde L_b = p_0 \dot x_0 - p_i \dot x_i - \frac{1}{2} e (p_0^2 -p_i^2 - m^2) 
- \frac{1}{2} (\dot e)^2 - i \dot {\bar c} \dot c$. These can be
derived from their off-shell versions
(5.2), (5.3) and (5.1), respectively, by the substitution of
the equation of motion $b = - \dot e$ emerging from the Lagrangian (5.1).}
are generated by the conserved and
off-shell nilpotent ($Q_{(a)b}^2 = 0$) (anti-)BRST charges $Q_{(a)b}$
as given below:
$$
\begin{array}{lcl} 
{\displaystyle
Q_b = b \dot c + \frac{c}{2} (p_0^2 - p_i^2 - m^2) \equiv b \dot c - \dot b c,
\quad
Q_{ab} = b \dot {\bar c} + \frac{\bar c}{2} (p_0^2 - p_i^2 - m^2) 
\equiv b \dot {\bar c} - \dot b \bar c},
\end{array} \eqno(5.4)
$$
because $s_{(a)b} \phi = - i [\phi, Q_{(a)b}]_{\pm}$ is true for 
the generic field $\phi = x_0, x_i, p_0, p_i, e$ of the theory. 
The subscripts $(+)-$ on the square bracket
correspond to the (anti-)commutators for the generic field $\phi$ being
(fermionic)bosonic in nature.
The physicality criteria $Q_{(a)b} |phys> = 0$, on the physical states of the
total Hilbert space, imply that the real physical
states $|phys>$ are annihilated by the operator form of
the first-class constraints
$\Pi_e = b$ and $\dot b = - (1/2) (p_0^2 - p_i^2 - m^2)$. In other 
words, the conditions
$\Pi_e |phys> = 0$ and $(p_0^2 - p_i^2 - m^2) |phys> = 0$ are
consistent with the Dirac's prescription for the quantization of theories,
endowed with the first-class constraints. In physical terms, the primary
constraint condition $\Pi_e |phys> = 0 (\Rightarrow b |phys> = 0)$ remains
intact with respect to the time evolution of the system because its 
time derivative
(i.e. $\dot b |phys> = 0 \Rightarrow (p^2 - m^2) |phys> = 0$) corresponds
to the annihilation of the physical states by the secondary constraint
$(p^2 - m^2 = 0)$.

The nilpotency ($Q_b^2 = 0$) property of the conserved BRST charge $Q_b$
and the physicality criteria ($Q_b |phys> = 0$) are the two key ingredients
for the definition of the BRST cohomology. In fact, two physical states
$|phys>^\prime (= |phys> + Q_b |\chi>)$ and $|phys>$ are said to belong
to the same cohomology class (i.e. $Q_b |phys>^\prime = 0 \Leftrightarrow
Q_b |phys> = 0$) w.r.t. the conserved and nilpotent BRST 
charge $Q_b$ if they differ by a BRST exact (i.e. $Q_b |\chi>$)
state where $|\chi>$ is any arbitrary non-null state of the
quantum Hilbert space. Since the BRST transformations $s_b$
(i.e. $s_b \phi = - i [\phi, Q_b ]_{\pm}$ for the generic field $\phi$) imbibe
the nilpotency property of $Q_b$, the cohomologically equivalent
transformations can be defined in terms of the nilpotent $s_b^2 = 0$
BRST transformations. For instance, the BRST transformed spacetime
variables in (5.2) can be re-expressed as
$$
\begin{array}{lcl} 
x_0 \to X_0 &=& x_0 + c\; p_0 \Rightarrow
x_0 \to X_0 = x_0 +  s_b\; [x_0],\nonumber\\
x_i \to X_i &=& x_i + c\; p_i \Rightarrow
x_i \to X_i = x_i +  s_b\; [x_i],
\end{array} \eqno(5.5)
$$
which show that the untransformed spacetime physical variables $(x_i, x_0)$
and the transformed spacetime variables $(X_i, X_0)$
belong to the  same cohomology class w.r.t. the nilpotent transformations
$s_b$ as they differ (with each-other) by a BRST exact transformation. 
It should be noted that the above transformations do {\it not}
lead to any NC in the spacetime structure because the non-trivial
brackets (i.e. $\{X_0, X_i \}_{(PB)} = 0, \{X_i, X_j \}_{(PB)} = 0$), in the
transformed frames {\it and} the corresponding brackets 
(i.e. $\{x_\mu, x_\nu \}_{(PB)} = 0$) 
in the untransformed frames, are found to be zero. Let us focus on 
the specific gauge symmetry transformation (3.10) that has been obtained
from the non-standard gauge-type transformations (3.1) and (3.2). It is
straightforward to obtain the off-shell as well as the on-shell
BRST symmetry transformations corresponding to this specific gauge
transformation. All one has to do is
to replace the local gauge parameter $\xi (\tau)$ by an 
anticommuting number and the (anti-)ghost fields $(\bar c)c$
(dictated by the (anti-)BRST prescription).
The nilpotent (anti-)BRST transformations can be written in a similar
fashion as given in (5.2) and (5.3). In fact, in the language of the
BRST transformations corresponding to (3.10),
the transformations on the spacetime variables can 
explicitly be written as

$$
\begin{array}{lcl} 
x_0 \to X^{(sp)}_0 = x_0 + {\displaystyle \frac{m}{\surd 2}}\; c, \qquad
x_i \to X^{(sp)}_i = x_i + {\displaystyle \frac{m}{\surd 2}}\; \theta_{i0}\;c .
\end{array} \eqno(5.6)
$$
It is straightforward to check that the non-trivial Poisson-bracket
$\{X^{(sp)}_0 (\tau), X^{(sp)}_i (\tau) \}_{(PB)} = - 2 c(\tau) \theta_{i0} 
\equiv \Theta_{0i} (\tau)$
is non-zero leading to the NC in the spacetime structure. In the above
computation, the NC of the mass parameter $m$ has been taken into account
and the brackets: $\{ x_0, m \}_{(PB)} = \surd 2, \{ m, x_i \}_{(PB)}
= + \surd 2\;\theta_{i0}$, emerging from equation (3.12), have been exploited.
This demonstrates that the NC of the transformations
(5.6) and the commutativity of the transformations (5.5) are different
aspects of the gauge symmetry transformations. This agrees with the
discussions about such an equivalence provided in [19] where the
language of the Dirac bracket formalism, for different
choices of the gauge-conditions,  has been exploited for the discussion
of reparametrization invariant theories.

It is obvious that the transformations (3.10) have been obtained
primarily from the basic non-standard gauge type of transformations
(3.1). It would have been pretty difficult to guess these transformations
(i.e. (3.10)) from the usual gauge transformations (2.5).
To have a closer look at the NC and commutativity discussed above,
let us concentrate on the basic transformations (3.1) and argue
in the language of the BRST cohomology. With the identification
$\zeta (\tau) = \xi (\tau)$ and the application of the BRST prescription,
the transformations (3.1) can be written in the language
of the BRST transformations in (5.2), as
$$
\begin{array}{lcl} 
&&x_0 \to X_0 = x_0 + \theta_{0i}\; c\; p_i 
\equiv  x_0 + s_b\; [\theta_{0i} x_i], \nonumber\\
&&x_i \to X_i = x_i + \theta_{i0}\; c\; p_0
\equiv  x_i + s_b\; [\theta_{i0} x_0]. 
\end{array} \eqno(5.7)
$$
The above transformations lead to the NC in the spacetime structure because
the non-trivial bracket (i.e. $\{X_0, X_i\}_{(PB)} = - 2 c \theta_{0i}$) is
non-zero. Here
we have used the basic canonical brackets $\{x_0, p_0\}_{(PB)} = 1,
\{x_i, p_j\}_{(PB)} = \delta_{ij}$, etc., and as before, the
antisymmetric (i.e. $\theta_{0i} = - \theta_{i0}$) NC parameter
is treated as a constant tensor. It is elementary to note that, 
once again, the
spacetime untransformed variables $(x_i, x_0)$ and the transformed
variables $(X_0, X_i)$ belong to the same cohomology class w.r.t.
the BRST transformations $s_b$. Thus, it is clear that the NC and
commutativity for the reparametrization invariant model for the
free massive relativistic particle belong to the same cohomology class
w.r.t. the nilpotent BRST transformation $s_b$. All the above  arguments
could be repeated with the nilpotent
anti-BRST transformations $s_{ab}$ 
(and corresponding conserved and nilpotent charge $Q_{ab}$), too,
to demonstrate the above equivalence.\\

\noindent
{\bf 6 Conclusions}\\

\noindent
The emphasis, in our present endeavour, is laid on the continuous
symmetry properties of the Lagrangian(s) for the physical system of
a free massive relativistic particle and their role in the description of the
commutativity and NC of the spacetime structure. In particular, the
reparametrization, standard gauge- and non-standard gauge-type symmetries play
very important roles in our whole discussion. To be specific, a set
of non-standard gauge-type transformations (cf. (3.1)) for the
space and time variables has been taken
into account to demonstrate the existence of a
NC in the spacetime structure. This NC primarily exists in the
transformed frames. One of the interesting points in our discussion
is the fact that these non-standard transformations, leading to a NC,
can be guessed from the corresponding
standard gauge transformations which
lead to the {\it commutative} transformed frames. In particular, to obtain
the non-standard symmetry transformations 
(e.g. (3.1)) from the standard gauge symmetry transformations 
(2.6), the infinitesimal gauge increments in the time and space variables are
to be exchanged with each-other. For the present model under consideration,
this trick has been performed through the introduction of the
noncommutative parameter $\theta_{0i}$ (cf. (3.1))
which appears in the context of the
time-space NC (i.e. $\{X_0, X_i \}_{(PB)} = - 2 \zeta \theta_{0i}$). 
This technique works here in our present endeavour because
the model under consideration
is nothing but the generalization of our earlier work
on the reparametrization invariant 
toy model of a non-relativistic particle [22].
The basis for such a trick comes from the fundamentals of the
BRST cohomology which is discussed, in detail, in Section 5. 
At the moment, it appears to us that the above
procedure will work for all the reparametrization invariant theories
where the equivalence between the commutativity and NC has been established
through Dirac bracket procedure for the specific choices of the 
gauge conditions that lead to the existence of NC and commutativity 
(see, e.g., [19]).

One of the most interesting features of our present investigation
on the reparametrization invariant model of a free massive
relativistic particle
is the existence of a very specific symmetry transformation (3.10)
which can be looked upon in {\it two} distinctly different ways. First,
it is a particular case of the usual gauge transformations (2.5)
(corresponding to the {\it commutative} spacetime) when we put
$p_0 = (m/\surd 2)$ and $p_i = \theta_{i0} (m/\surd 2)$ in the 
expression for the infinitesimal increments in the
time and space variables.
Second, as discussed in detail in Section 3, this transformation
can be obtained from the non-standard gauge-type transformations
(3.1) and (3.2) (corresponding to the {\it noncommutative} spacetime)
if we demand the consistency among (i) the basic non-standard
transformations for the time and space variables, (ii) the
expressions for the canonical momenta derived from various
equivalent Lagrangians for the system, and (iii) the equations
of motion derived from the above Lagrangians. The outcome of these
requirements leads (i) to enforce the mass parameter of the model
to be noncommutative (cf. (3.12)) with both the time and space
variables, and (ii) to restrict the antisymmetric
(i.e. $\theta_{0i} = - \theta_{i0}$) noncommutative parameter
$\theta_{0i}$ to obey $\theta_{0i} \theta_{0j} = - \delta_{ij}$.
The NC of the mass parameter $m$, in the context of
noncommutative geometry, is {\it not} a speculative and strange idea.
For the free motion of the (non-)relativistic (super-)particles
on a ``quantum'' (super) world-line, it has been shown that the mass 
parameter could become noncommutative [31, 25-28] in the framework of quantum 
groups where the NC of the mass parameter
is altogether of a different variety than our present discussion
(in which the NC is of a kind given by (3.12)).

A noteworthy point in our discussion is a minor deformation
of the Poincar{\'e} algebra by a constant factor and a major
deformation in the algebra between the angular momentum generator
$M_{\mu\nu}$ and the spacetime variable $x_\mu$ 
due to the presence of a NC in the spacetime structure (see, e.g., Section 4). 
In particular,
the Poisson brackets that include the boost generator $M_{0i}$
pick up an additional factor of two (cf. (4.4),(4.7)) because of the NC
of the mass parameter with both the space and time variables (cf. (3.12)).
This observation should be contrasted with (i) the results of
[19] where the Dirac brackets, corresponding to the Poincar{\'e}
algebra, do {\it not} get any contribution from the NC of spacetime
because the noncommutative parameter $\theta_{0i}$ does not appear
in the algebra, and
(ii) the results of [18] where the angular momentum generator
itself is modified by an extra term for the closure of the algebra.
Furthermore, in [19], the NC parameter $\theta_{0i}$ appears
only in the transformed frames for the Dirac algebra between
the angular momentum generator $M_{\mu\nu}$ and the spacetime
variable $x_\mu$ but it (i.e. $\theta_{0i}$) does not appear in the
untransformed frames exactly for the same Dirac algebra. 
The above observations should be contrasted with
our discussion where (i) the Poisson bracket algebra remains
{\it form-invariant} in both the (un-)transformed frames, and (ii)
the noncommutative parameter $\theta_{0i}$ appears in both the 
transformed as well as untransformed frames (cf. (4.3)).

To obtain a time-space NC in the spacetime structure, our approach is
a general one. This claim is valid 
 in the sense that it can be applied to any arbitrary reparametrization
invariant theories (because, as stressed earlier,
the basis for our trick comes from the basics
of the BRST cohomology). Our method can be applied now to the reparametrization
invariant model of a free relativistic superparticle that has already been
discussed in the framework of the (super) quantum  groups 
$GL_{\surd q}(1|1)$ and $GL_q (2)$ [26]. It would also be a nice endeavour 
to find a possible connection between these two different approaches. 
The generalization
of our present work to the study of a reparametrization invariant
model of an interacting relativistic particle, with 
the electromagnetic field in the background, is yet another direction that
could be pursued in the future. These are some of the issues 
that are under investigation and our results would be reported elsewhere [32].

\end{document}